\documentstyle[11pt,newpasp,twoside,epsf]{article} 
\def\edcomment#1{\iffalse\marginpar{\raggedright\sl#1\/}\else\relax\fi} 
\reversemarginpar 

\newcommand{\kms}{\,km\,s$^{-1}$}
\def\edcomment#1{\iffalse\marginpar{\raggedright\sl#1\/}\else\relax\fi}
\reversemarginpar

\begin{document}
\title{Reconstruction of the Zone of Avoidance}
\author{Yehuda Hoffman}
\affil{Racah Inst. of Physics, Hebrew University, Jerusalem 91904,
        Israel}

\begin{abstract}
The uncovering of the large-scale structure that is hidden in the Zone
of Avoidance is essential to our quest for the understanding of the
dynamics of the nearby universe.  Dedicated and sophisticated
observations of the Zone of Avoidance have to be matched by equally
sophisticated theoretical tools for the reconstruction of the
large-scale structure from sparse, noisy and very incomplete data. A general
approach for the estimation of cosmological parameters and the
reconstruction of the density and velocity fields is presented here,
based on maximum likelihood analysis, Wiener filtering and constrained
realizations.  Radial velocity surveys and the IRAS 1.2\,Jy redshift
survey are used to reconstruct the Zone of Avoidance.
\end{abstract}

\section{Introduction}

The galactic Zone of Avoidance (ZOA) presents an obstacle to our
efforts to understand the dynamics of the `local' universe. Namely,
the Milky Way prevents a full-sky mapping of the light, and therefore
mass, distribution that induces the local velocity field. Thus, a prime
motivation of the astronomical study of the ZOA is to survey the
galaxy distribution in the regions obscured by the Milky Way. These
efforts have indeed uncovered a very rich structure in the ZOA, in
particular in the region of the Great Attractor (GA; Kraan-Korteweg \&
Lahav 2000, and references therein).

In spite of all the efforts and successes in unveiling the ZOA, the
fact remains that the ZOA is a region more difficult to observe than
other parts of the sky.  It follows that the reconstruction of the
large-scale structure (LSS) in the ZOA should rely more on sophisticated
statistical methods than other fields of the sky.  Here, we
shall report on a Bayesian approach to the reconstruction of the LSS
from incomplete, sparse  and noisy data sets, such as peculiar velocities and
redshift surveys.  Most of the analysis centers on the recovery of the
LSS from noisy surveys of radial velocities with very incomplete sky
coverage , with some analysis of the IRAS 1.2 Jy redshift survey that
is presented as well.

The main focus of the present paper is on the general methodology with
an emphasize on the tools and methods relevant to the reconstruction
of the LSS. The method is applied to the MARK III (Willick et
al. 1995), SFI (da Costa et al. 1996) and the ENEAR (da Costa
et al. 2000) surveys of the radial velocities, as well as the
IRAS 1.2\,Jy redshift survey.
Most of the work presented here has been done in
collaboration mostly with S. Zaroubi and the cosmographical
implications of this work are presented in Zaroubi (these
proceedings).  Here, the
general Bayesian methodology is presented in \S 2.  The cosmological
parameters estimated from radial velocities surveys are summarized in
\S 3.  The LSS and the ZOA that is reconstructed from radial velocity
catalogs and the IRAS 1.2\,Jy redshift survey are presented in \S 4.  A
new algorithm of recovering the tidal field from the full velocity
field enables the studying of the LSS on scales beyond those that are
directly probed by the data (\S 5).  A technique of simulating the
dynamical role of the ZOA based on N-body simulations that are
constrained by the observed LSS is described in \S 6.  A general
discussion concludes this review.

\section{Theory: Parameter Estimation and Wiener Filtering}

The general methodology of analyzing surveys of the LSS, such as
redshift and radial velocity surveys, was extensively presented in
Zaroubi et al. (1995).  The main result  of the paper is that for
Gaussian random fields an optimal estimation of the cosmological
parameters is provided by a maximum likelihood analysis.  The
underlying density and velocity fields are reconstructed by a Wiener
filter (WF) applied to the data.  Monte Carlo realizations of the
scatter of the residual from the WF fields are done by means of
constrained realizations (Hoffman \& Ribak 1991; CR).  A brief
summary is presented here.

Consider a data set of radial velocities $\{ u_{i}\}_{i=1,\ldots,N}$,
where
\begin{equation}
u_{i} = {\bf  v}
({\bf r}_{i})  \cdot \hat {\bf r}_{i}
+\epsilon_{i},
	\label{eq:ui}
\end{equation}
${\bf  v}$ is the three dimensional velocity, ${\bf r}_i$ is the
position of the i-th data point  and $\epsilon_{i}$ is the statistical
error associated with the i-th radial velocity. The assumption made
here is of   a   cosmological model that
describes the data well, that systematic errors have been properly dealt
with and that the statistical errors are well understood.
The data auto-covariance matrix  is then  written as:
\begin{equation}
R_{ij} \equiv
\Bigl < u_i u_j  \Bigr > =   \hat {\bf r}_j \Bigl < {\bf  v}
({\bf r}_i) {\bf  v} ({\bf r}_j)  \Bigr >
   \hat {\bf r}_j   + \sigma{^2_{ij}}.
\label{eq:Rij}
\end{equation}
(Here $\Bigl <  \ldots  \Bigr >$ denotes an ensemble average.)
The last term is the error covariance matrix.
The velocity covariance tensor is calculated within the linear theory.

The likelihood  of the $N$ data points given a model is,
\begin{equation}
{\cal L} = [ (2\pi)^N \det(R_{ij})]^{-1/2}
  \exp\left( -{1\over 2}\sum_{i,j}^N {u_i R_{ij}^{-1} u_j}\right).
\label{eq:like}
\end{equation}
An unbiased estimation of the cosmological parameters is given by a
maximum likelihood analysis which finds the most probable parameters given the
data and within a model space.


The general application of the WF/CR method to the reconstruction of
LSS is described in Zaroubi et al. (1995),
where the theoretical foundation is discussed in relation to other
methods of estimation, such as Maximum Entropy.
The specific application of the WF/CR method to peculiar velocity data
sets has been presented in Zaroubi et al. (1999). Here we
provide only a
brief description of the WF/CR method and the interested reader is
referred to the last two references  for more details.

We assume that the peculiar velocity field ${\bf v}({\bf r})$ and the
density fluctuation field $\delta({\bf r})$ are related via the linear
gravitational-instability theory.  Under the assumption of a specific
theoretical prior for the power spectrum $P(k)$ of the underlying
density field, one can write the WF minimum-variance estimator of the
fields as
\begin{equation}
{\bf v}^{WF}({\bf r}) =
\Bigl < {\bf v}({\bf r}) u_i \Bigr >  \Bigl < u_i u_j \Bigr >
^{-1}        u_j
\label{eq:WFv}
\end{equation}
and
\begin{equation}
\delta^{WF}({\bf r}) = \Bigl < \delta({\bf r}) u_i \Bigr >  \Bigl < u_i u_j
\Bigr > ^{-1}    u_j  .
\label{eq:WFd}
\end{equation}

A well known problem of the WF is that it attenuates the estimator to
zero in regions where the noise dominates. The reconstructed mean
field is thus statistically inhomogeneous.  In order to recover
statistical homogeneity we produce constrained realizations (CR), in
which random realizations of the residual from the mean are generated
such that they are statistically consistent both with the data and the
{\it prior} model (Hoffman \& Ribak 1991).
In regions dominated by good quality data, the CRs are dominated by
the data, while in the limit of no data the realizations are
practically unconstrained.

\section{Estimated Parameters}

A maximum likelihood analysis has been performed on the MARK III
(Zaroubi et al. 1997), SFI (Freudling et al. 1999) and
ENEAR (Zaroubi et al. 2000) surveys of radial velocities.
Considering the constraints set on Hubble's constant (expressed by
$H_{0}=65 h_{65}$\kms\,Mpc$^{-1}$) and the total matter density parameter
($\Omega_{0}$), the data does not constrain each parameter seperately
but rather a certain non-linear combination of the two.  For the
COBE-normalized flat $\Lambda$CDM models, the most probable parameters
are given by:
\begin{equation}
\Omega_{0} h_{65}^{1.3} = 0.56 \pm 0.14 \ \ \ {\rm MARK\ III},
\end{equation}
\begin{equation}
\Omega_{0} h_{65}^{1.3} = 0.52 \pm 0.10 \ \ \ {\rm  SFI,}\ \ \ \ \ \ \  \
\end{equation}
\begin{equation}
\Omega_{0} h_{65}^{1.3} = 0.66 \pm 0.13 \ \ \ \ {\rm  ENEAR.}\ \ \
\end{equation}
All quoted uncertainties refer to   $3\sigma$ confidence levels.

For observables that are drawn from a Gaussian random field, a
goodness-of-fit criterion for a model to be consistent with the data is
that the total $\chi^{2}$ per degree of freedom should be close to
unity.  Indeed, for all velocity surveys considered, the most likely
models obey this criterion.  However, given the complex nature of the
data (noise, selection etc.)  the goodness-of-fit criterion has been
extended and performed on a mode-by-mode basis (Hoffman \& Zaroubi
2000).  The idea is to use the representation of the data by the
eigenmodes of the data auto-covariance matrix.  The analysis reveals
that for all surveys there is a systematic trend of the $\chi^{2}$ to
increase with the mode number (where the modes are sorted by
decreasing value of the eigenvalues).  This suggests a systematic
inconsistency of the assumed theoretical model and/or error analysis
with the data.  This trend is statistically very significant in the
case of the MARK III and SFI surveys, and less so in the case of
ENEAR.

\section{Reconstruction of the Large-scale Structure}

The WF/CR algorithm has been applied both to MARK III (Zaroubi
et al. 1999), SFI (Hoffman \& Zaroubi, unpublished) and the ENEAR
(Zaroubi et al. 2000) surveys. The LSS uncovered
by the velocity surveys is shown in Fig. 1, where the
density and velocity evaluated at the Supergalactic plane are shown.
A view of that structure in galactic Aitoff projection, focusing on
the ZOA, is presented in Zaroubi (these proceedings). A comparison of the
structure revealed by the three structure shows a gross agreement
with respect to the main `players' in the local dynamics, however the
fine structure differs. Inspection of the large-scale velocity field
reveals that in all cases the flow is dominated by two main
structures, the  GA and the Perseus-Pisces (PP) supercluster.
However, the MARK III exhibits a long range coherent component of the
velocity field which is clearly induced by distant structures, not
directly probed by the data. This component is smaller in the SFI case
and is missing altogether from the ENEAR velocity field.

\begin{figure}
\vspace{-1cm}
\plotone{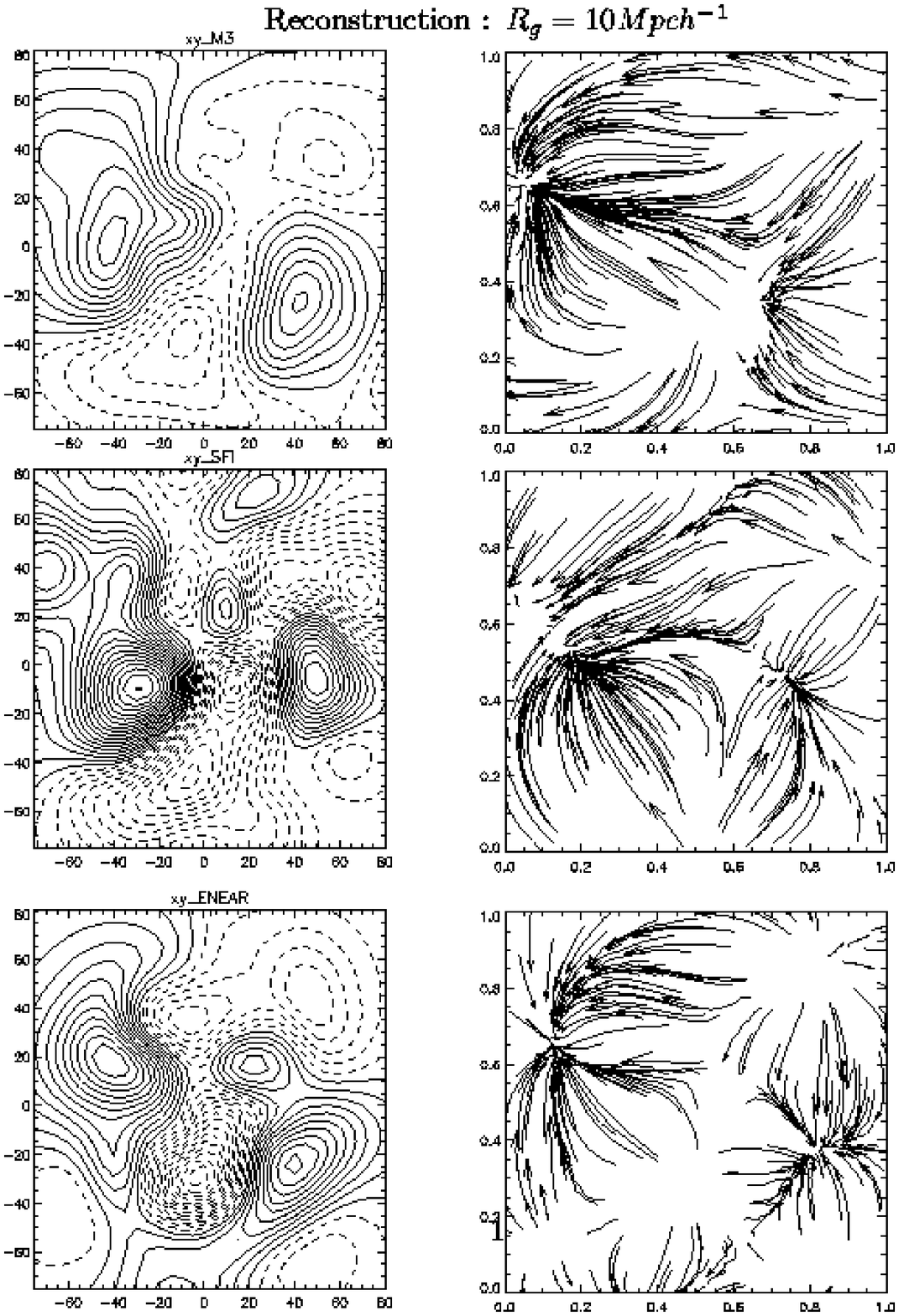}
\caption{ Reconstruction of the large scale density (left panels) and
velocity (right panels) fields along the Supergalactic Plane
from the MARK III  (upper), SFI (middle) and ENEAR (bottom panels)
surveys. Contour spacing is $0.1$. The velocity field is represented
by stream lines. (Distances are given in units of $h^{-1}$\,Mpc,
where $h$ is Hubble's constant in units of 100\kms\,Mpc$^{-1}$.)}
\label{fig:wf_sgp}
\end{figure}

The WF/CR has been applied to the IRAS 1.2\,Jy redshift survey (Bistolas
\& Hoffman 1998, Bistolas 1999).  A thorough analysis of the LSS
inferred by this survey is beyond the scope of the present paper.
Only the ZOA reconstruction is presented here.  A study of the
cosmography unveiled in the ZOA is to be reported elsewhere (Bistolas,
Kraan-Korteweg, \& Hoffman, in preparation). Here, the ZOA is shown in
Aitoff projections of spherical slices ranging from $1500$ to
$8000$\kms\ (Figs. 2 and 3).

\begin{figure}
\plotone{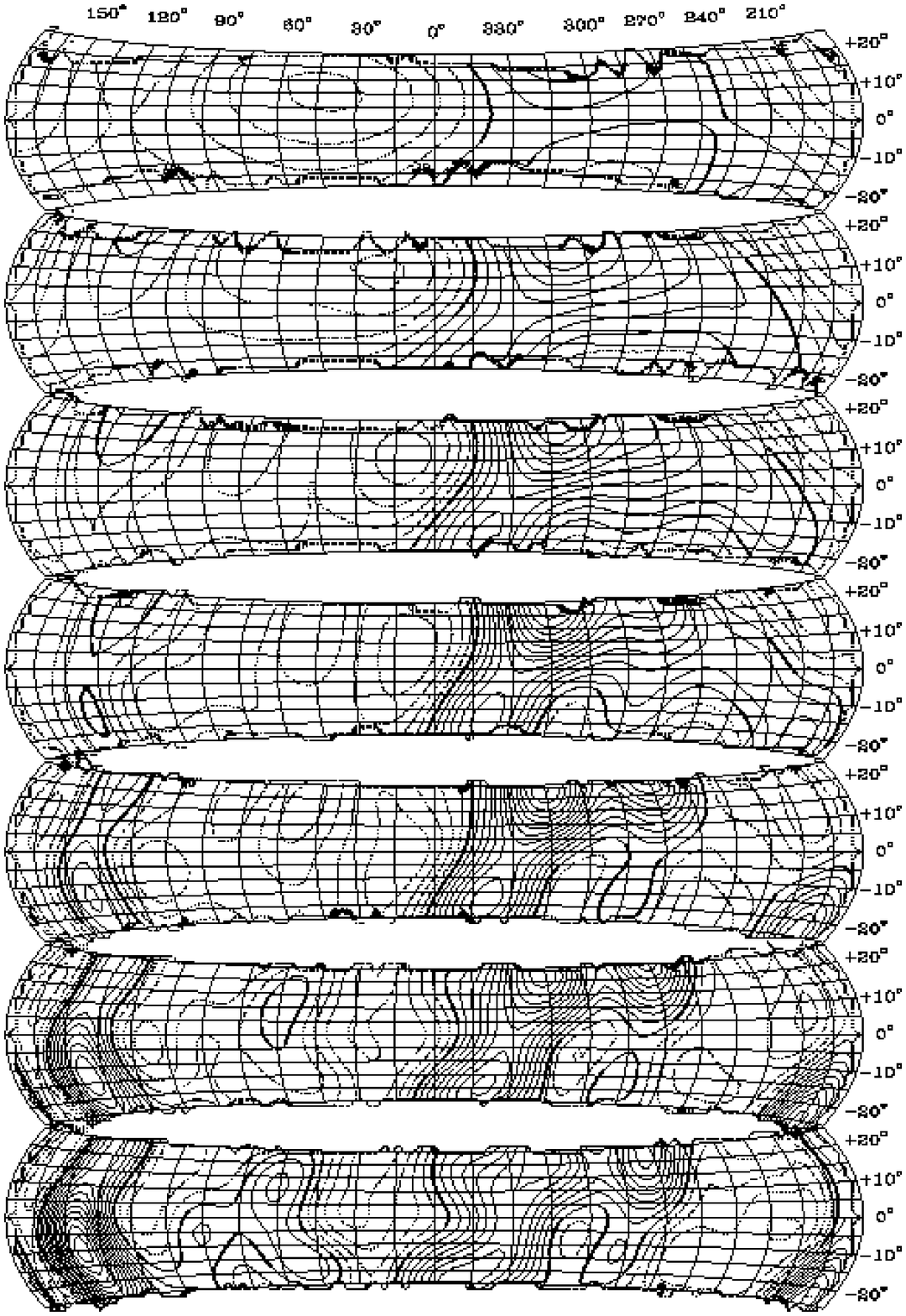}
\caption{Aitoff plots in galactic coordinates of  the density field
reconstructed from the IRAS 1.2\,Jy data. The
 $ \vert b \vert <  25\deg$ strip
is shown here. From top to bottom: $1500, 2000, 2500,3000, 4000,
4500$\kms\ shells. The contour spacing is $0.2$ in $\delta$.  The field
is smoothed with a Gaussian kernel of $5 h^{-1}$\,Mpc.}
\label{fig:iras_aitoffa}
\end{figure}

\begin{figure}
\plotone{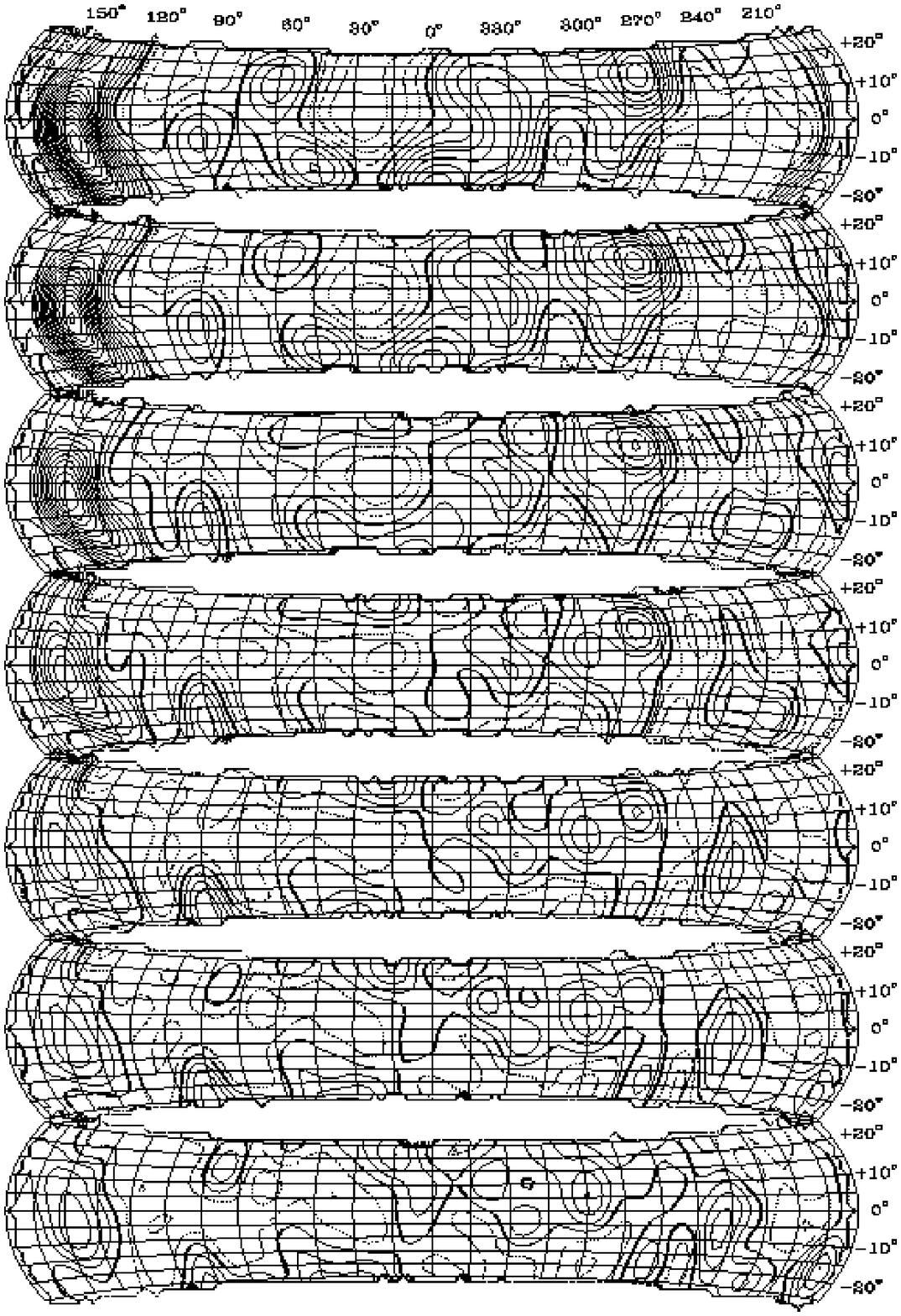}
\caption{Same as Fig. 2 with shells of 5000 to
8000\kms, in intervals of 500\kms.}
\label{fig:iras_aitoffb}
\end{figure}

\section{Tidal Field}

The velocity field can be decomposed into two components, one which is
induced by the local mass distribution and a tidal component due the
external mass.  Here, we follow the procedure suggested by Hoffman
(1998a,b) and more recently by Hoffman et al. (2000).  The key idea is
to solve for the particular solution of the Poisson equation with
respect to the WF density field within a given region and zero padding
outside.  This yields the velocity field induced locally, hereafter
the divergent field.  The residual of the full velocity field is the
the tidal field.  Figure 4 shows the decomposition
applied to the MARK III survey, where the local volume is a sphere of
6000\kms\ centered on the Local Group.  The plots show the full
velocity field, the divergent and tidal components.  To further
understand the nature of the tidal field, its bulk velocity component
has been subtracted and the residual is shown as well.  This residual
is clearly dominated by a quadrupole component.  In principle, the
analysis of this residual field can shed light on the exterior mass
distribution.

Indeed, the full velocity field recovers
the CMB dipole velocity, half of which is contributed by the tidal
field that is induced by structures outside of the 6000\kms\ sphere.
The tidal field is decomposed into a dipole and quadrupole moments,
namely a bulk velocity and traceless shear tensor.  The directions of
the bulk velocity and the eigenvectors of the shear tensor are shown
in Aitoff projection in galactic $(\ell,b)$ coordinates (Fig.~5).
The figure shows that both the bulk velocity and
the dilational eigenmode lie in, or close to, the ZOA. The tidal field
has a bulk velocity with an amplitude of $\approx 300$\kms\ in the
direction of (galactic) $(\ell,b)\approx (300\deg,15\deg)$.   It follows that
the deep ZOA (on scales larger than 6000\kms) induces a substantial
part of the local dynamics.

\begin{figure}
\plotfiddle{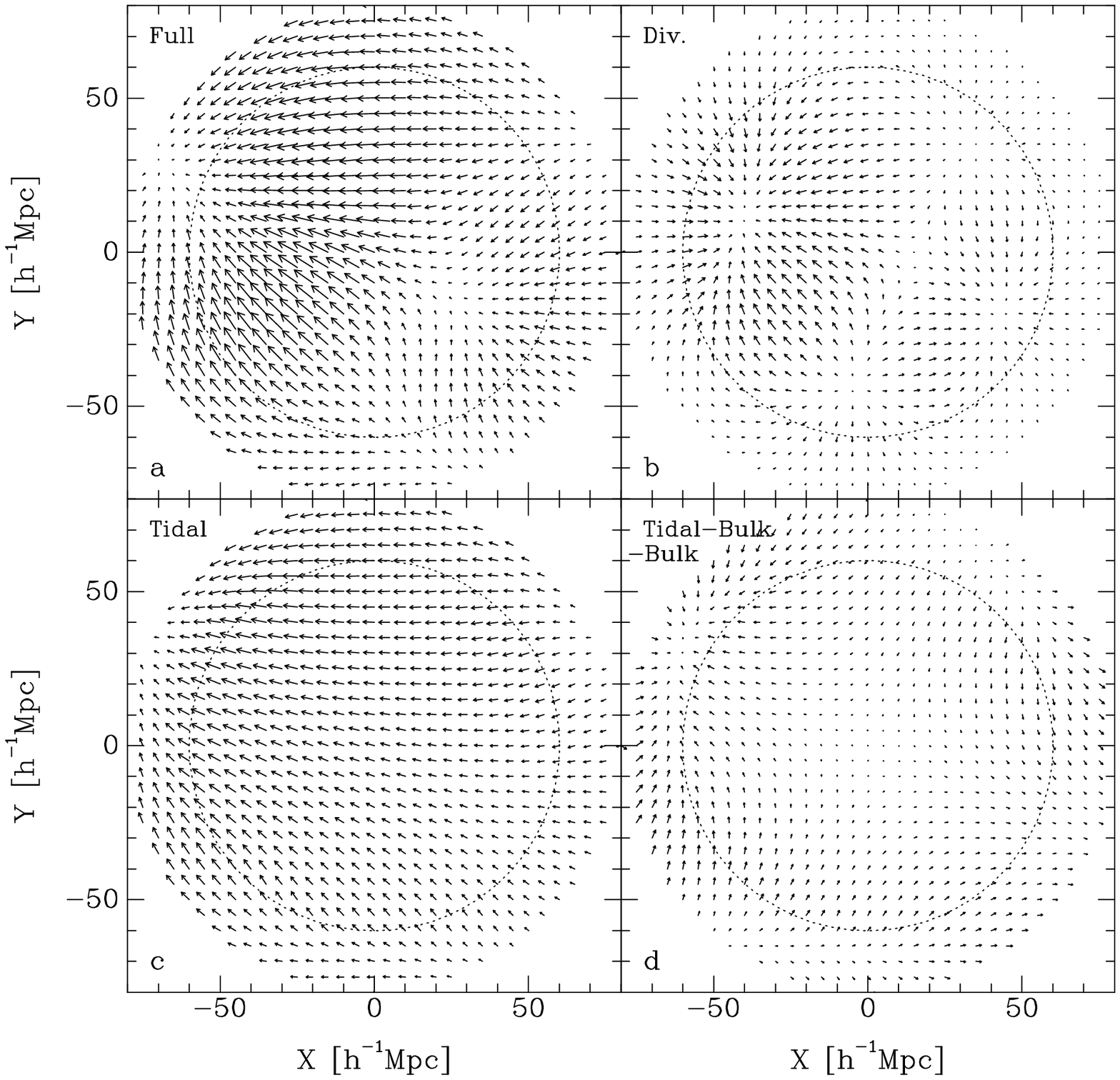}{150truemm}{0}{69}{69}{-210}{-20}
\caption{Tidal field decomposition of the
velocity field obtained by WF of the MARK III survey: full velocity
field (upper left), divergent (upper right) and the tidal field (lower
left) are shown. The lower right panel shows the residual of the tidal
field from its bulk velocity moment. (Distances are given in units of
$h^{-1}$\,Mpc.)
}
\label{fig:tidal-field}
\end{figure}

\begin{figure}
\plotone{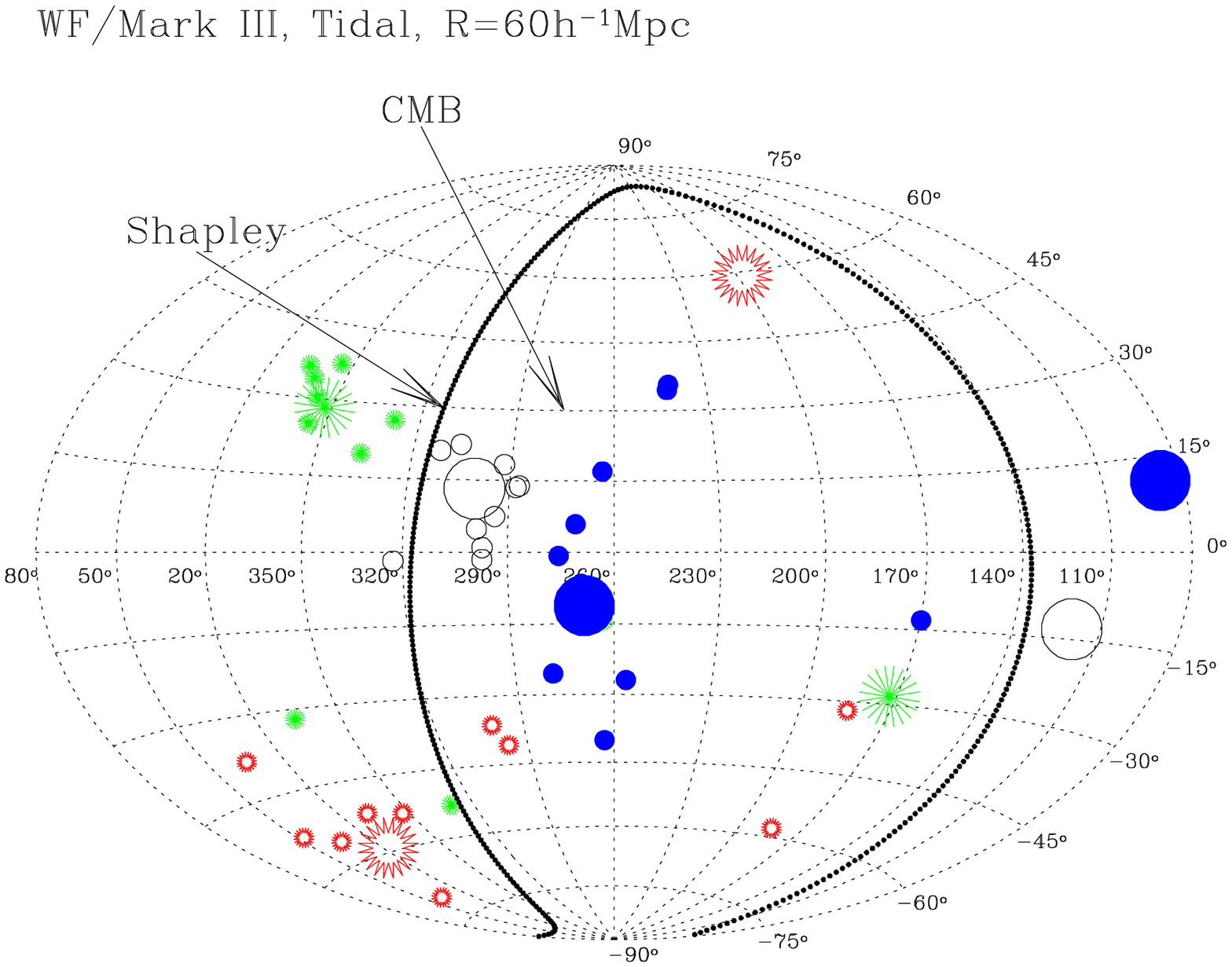}
\caption{Directions in the sky, in galactic coordinates, of the bulk velocity
(open circles), and three eigenvectors of the shear: dilation (skeletals),
compression (starred), and the middle-eigenvalue vector (solid).
The large symbols are for the WF reconstruction, and small symbols
are for 10 CRs. The anti-pole of the WF eigenvectors is presented as
well.
}
\label{fig:tidal_aitoff}
\end{figure}

\section{Simulations of the Zone of Avoidance}

The WF/CR is formulated within the linear theory of the gravitational
instability, where the deviations from a pure Friedmann model
constitute a Gaussian random field.  However, as these deviations
evolve away from the linear regime, the use of linear tools is of very
limited value.  This algorithm can be extended to provide quasi
non-linear CRs.  The basic idea is to use constraints that are drawn
from data that is linear or quasi-linear, to generate linear CRs given
a {\it prior} model, take it backwards in time and feed these
realizations as initial conditions to N-body simulations.  The final
configuration is obtained by solving exactly the non-linear dynamics
subject to the   boundary conditions imposed by the actual universe.
This provides   realistic N-body simulations that reproduce the
observed universe, e.g. the GA, PP and the Local Void with the
right location and amplitude. Such simulation are essential for
non-linear modeling of the dynamical role of the ZOA in surveys of
the LSS.

An application of this was done by Bistolas \& Hoffman
(1998) who used the IRAS 1.2 redshift survey as the constraining data.
The resulting structure reproduces the observed galaxy
distribution out to a distance of $\approx 5000$\kms\
(Fig.  6).  This procedure has been applied also
to the MARK III data (van de Weygaert \& Hoffman 2000; Klypin, van
 de Weygaert, \& Hoffman, in preparation).

\begin{figure}
\plotone{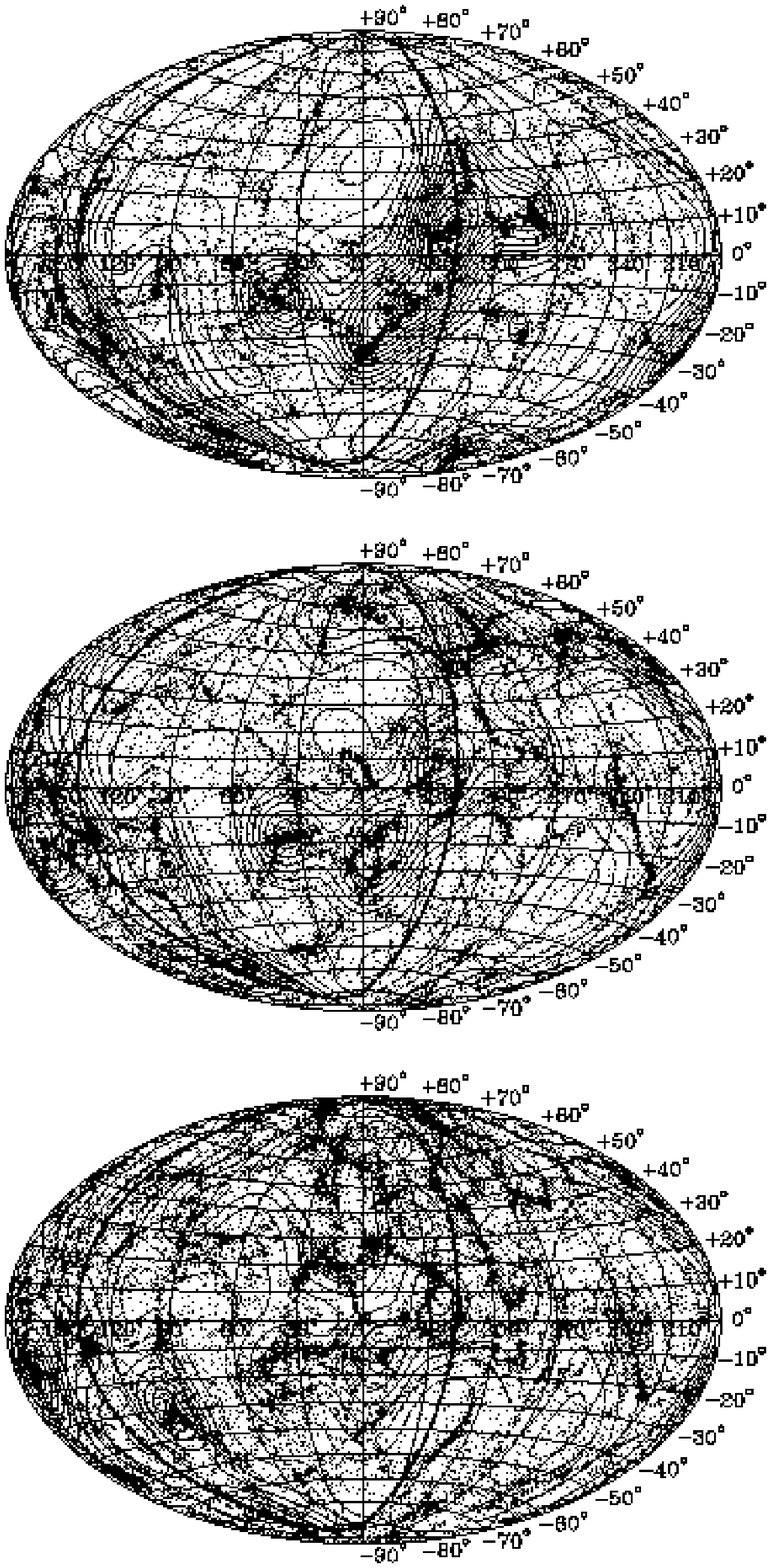}
\caption{Aitoff plots in galactic coordinates of the non-linear
density field reconstructed from the IRAS 1.2\,Jy data. The $R=20,30,
40 h^{-1}$\,Mpc shells are shown. The thick curves marks the
Supergalactic plane and the contour line represent the $5 h^{-1}$\,Mpc
smoothed density field. The particles represent a mock volume-limited
IRAS survey, within a $10 h^{-1}$\,Mpc spherical slice centered on
that shell.}
\label{fig:iras_nlcr-20}
\end{figure}


\section{Discussion}

The recovery of the LSS in the ZOA poses a challenge to our
understanding of the dynamics of our `local' universe.  Direct
observational efforts of unveiling the ZOA, sophisticated as they might be,
need to be complemented by elaborate statistical method for
the reconstruction of the LSS that is hidden in the ZOA. Within the
standard cosmological model in which structure emerges from a
primordial Gaussian perturbation field {\it via}  gravitational
instability, the Bayesian approach presented here provides   the
optimal tools for estimating the cosmological parameters and the
reconstruction of the LSS.

The algorithm presented here consists of the following four steps.
First, given a data set, the cosmological (and other) parameters are
estimated by maximum likelihood analysis over a range of models.
Then, the consistency of the most probable model with the data is
analyzed by the differential $\chi^{2}$ analysis.  The model that is
found by the maximum likelihood and the goodness-of-fit analysis,
serves as the {\it prior} model for the WF/CR reconstruction.
Applying the WF to the data, one gets the mean density and velocity
fields, thus providing an estimator of the LSS. The CRs
constitute an ensemble of Monte Carlo simulations that are designed to
reproduce the imposed constraints, namely the observed radial velocity
and/or redshift surveys.  The scatter of the CRs around the WF field
provides a measure to the robustness of WF reconstruction.

A detailed study of the cosmography recovered from the radial velocity
surveys is given by Zaroubi (these proceedings).  Here, this has been extended to
study structure beyond the depth of these surveys.  This is done by
the tidal field decomposition of the velocity field.  The tidal field
induced by the structure beyond a sphere of radius 6000\kms\ has a
dipole component of 300\kms\ towards (galactic) $(\ell,b)\approx
(300\deg,15\deg)$. This direction implies that a substantial fraction
of the structure that induces this dipole is hidden in the ZOA. This
provides a further motivation for studying the ZOA.

\acknowledgments
The present review is based mostly on work done in
collaboration with S. Zaroubi.  A. Eldar is gratefully
acknowledged for his help with producing some of the figures.  I am
grateful to the Local Organizing Committee, and in
particular Ren\'ee Kraan-Korteweg, for organizing a very exciting
meeting.  This research has been partially supported by the Binational
Science Foundation (94-00185) and the Israel Science Foundation
(103/98).

\end{document}